 \documentclass[12pt,preprint]{aastex}

\newcommand{\mjb}{mJy~beam$^{-1}$}

\newcommand\kms{km~s$^{-1}$}

\shorttitle{OH (1720 MHz) Masers in SNR IC 443}
\shortauthors{Hoffman et al.}

\begin{document}

\title{The Sizes of OH (1720 MHz) Supernova Remnant Masers: \\ MERLIN and VLBA Observations of IC~443}

\author{I.\ M.\ Hoffman}
\affil{National Radio Astronomy Observatory, P.\ O.\ Box O, Socorro, NM, USA 87801}
\affil{Department of Physics and Astronomy, University of New Mexico, Albuquerque, NM, USA 87131}
\email{ihoffman@nrao.edu}

\author{W.\ M.\ Goss, C.\ L.\ Brogan, M.\ J Claussen}
\affil{National Radio Astronomy Observatory, P.\ O.\ Box O, Socorro, NM, USA 87801}

\and

\author{A.\ M.\ S.\ Richards}
\affil{MERLIN/VLBI National Facility, University of Manchester, Jodrell Bank Observatory, Macclesfield, Cheshire SK11 9DL, UK}

\begin{abstract}
MERLIN and VLBA observations of the 1720~MHz maser emission from the OH molecule in the supernova remnant IC~443 are presented.
Based on MERLIN data with a resolution of 160~mas, the deconvolved sizes of the maser sources are in the range 90 to 180~mas (135 to 270~AU).
The 12~mas resolution VLBA images show compact cores with sizes in the range 15 to 55~mas.
The maser brightness temperatures are $(2-34) \times 10^6$~K for the MERLIN sources and $(5-19) \times 10^8$~K for the VLBA cores, in agreement with theory.
Unlike the Zeeman Stokes $V$ profiles observed in other OH (1720~MHz) SNR masers, single-handed circular polarization line profiles are observed in IC~443 on all angular scales from 1000 to 10~mas resolution.
For one line component, the observed line width is 0.24$\pm0.07$~\kms, compared to an estimated Doppler width of 0.49~\kms.
This discrepancy in line widths can be accounted for if the maser emission arises from an elongated ellipsoidal region of masing gas.
\end{abstract}



\section{Introduction}

Maser emission from the OH (1720 MHz) satellite transition $({^2}{\Pi}_{3/2},J={3\over2},F=2{\rightarrow}1)$ in supernova remnants was first observed by Goss (1968) toward W28 and W44.
Study of OH (1720 MHz) masers in SNRs was then renewed by Frail, Goss, and Slysh (1994) toward the supernova remnant W28.
Subsequent studies and surveys (Yusef-Zadeh et al.\ 2000; Yusef-Zadeh et al.\ 1999; Koralesky et al.\ 1998; Green et al.\ 1997; Frail et al.\ 1996; Yusef-Zadeh et al.\ 1996) have observed $\sim 200$ Galactic supernova remnants and found that 22 remnants have associated OH (1720~MHz) maser emission.
Unlike their H{\sc ii} region counterparts ({\it e.g}.\ Gray, Doel, \& Field 1991), SNR OH (1720 MHz) masers are thought to be collisionally pumped ({\it e.g}.\  Elitzur 1976), and are never accompanied by masers from the other ground state OH transitions ({\it e.g}.\ Green et al.\ 1997; Frail et al.\ 1996).
Additionally, H{\sc ii} region OH masers are relatively small in size ($\sim 2$~mas) and are known to be variable ({\it e.g}.\ Masheder et al.\ 1994) while SNR OH (1720~MHz) masers are much larger ($\sim 200$~mas, \S 3.1) and have not been observed to vary.

Observations suggest that OH (1720~MHz) SNR masers occur where the shock front from a supernova explosion encounters a molecular cloud ({\it
e.g}.\ W28:  Frail, Goss, and Slysh 1994; Kes~78:  Koralesky et al.\ 1998; 3C391:  Frail et al.\ 1996; W44: Wootten 1977).
In such interactions, a C-type (non-dissociative) shock can produce the relatively rare conditions ($n \approx 10^5\,{\rm cm}^{-3}$, $T \approx 90$~K, $N_{\rm OH} \sim 10^{16-17}\,{\rm cm}^{-2}$) needed to produce collisionally pumped OH (1720 MHz) masers (Elitzur 1976; Wardle 1999; Lockett, Gauthier, \& Elitzur 1999, hereafter LGE; see also Draine \& McKee 1993).
Observationally, the VLBA-scale Zeeman magnetic field measurements from OH (1720~MHz) masers in W44, W28, and W51C (Claussen et al.\ 1999, hereafter C99; Brogan et al.\ 2002) verify the $C$-type shock predictions of the collisional pump.
The temperature, density, and OH column of the OH (1720~MHz) collisional pump are also well established by observation ({\it e.g}.\ Claussen et al.\ 1997, hereafter C97; C99).  

The SNR/molecular cloud interaction in IC~443 is especially well studied.
Many molecular observations have verified that the temperature ($T = 33-100$~K) and density ($n = (1-8)\times10^5$~cm$^{-3}$) are in accord with OH (1720~MHz) theory (Turner et al.\ 1992; Dickman et al.\ 1992; Ziurys, Snell \& Dickman 1989; van~Dishoeck et al.\ 1993; Cesarsky et al.\ 1999; Rho et al.\ 2001).
The OH (1720~MHz) masers occur in ``clump G'' of the shocked molecular interaction regions in IC~443 (DeNoyer 1979b; Huang, Dickman, \& Snell 1986) and, specifically, sub-clump G1 as resolved by submillimeter and interferometric observations (van~Dishoeck et al.\ 1993; Tauber et al.\ 1994).
Nonetheless, an important observational constraint that is still required is the transverse size of the region of OH inversion.  In addition, the fraction of this region that participates in the observed stimulated emission remains uncertain.

A fundamental problem in understanding OH (1720 MHz) SNR maser emission lies in quantifying the scatter-broadening of the imaged maser sizes.
The narrow-band maser emission does not permit multi-wavelength fitting of the scattering properties to the $\lambda^2$ wavelength dependence expected of ionized ISM scattering at radio frequencies.
It is not known if the resolved areas of OH (1720~MHz) maser emission reveal the true sizes of the emitting regions or are simply scatter-broadened images of smaller regions.
Many of the well-studied maser regions (such as W28 and W44) lie in directions of the Galaxy along which the scattering due to the ionized interstellar medium is expected to be prominent ({\it e.g}.\ Yusef-Zadeh et al.\ 1999; Kaspi et al.\ 1993; Taylor \& Cordes 1993; Cordes et al.\ 1991).
C99 have studied the OH (1720~MHz) masers in the supernova remnants W28 and W44 at 40~mas resolution with the VLBA and at 200~mas resolution with MERLIN.
They find angular sizes $30\ {\rm mas}<{\theta}<240\ {\rm mas}$ and argue that these sizes may be the intrinsic, unscattered sizes of the masers.
In contrast, some OH (1720~MHz) masers near the Galactic center (Yusef-Zadeh et al.\ 1996) have sizes expected to be dominated by angular scatter-broadening.

Claussen et al.\ (2002) have recently determined that the interstellar scattering of the W28 masers is minimal by studying the scattering properties of both an adjacent background pulsar and an extragalactic source.
In this paper, we seek to corroborate their results with an independent OH (1720~MHz) maser image size measurement.
In order to minimize the angular broadening effects caused by the interstellar medium, the OH (1720~MHz) masers in the supernova remnant IC~443 have been observed.
Both C99 and Frail, Goss, \& Slysh (1994) suggested that the IC~443 masers would be useful for this purpose, since this SNR lies toward the Galactic anticenter ($\ell = 189.1^\circ$, $b = 3.0^\circ$).
Indeed, sources in the direction of IC~443 are expected to show less than 1~mas of scatter-broadening (Lazio \& Cordes 1998a,b).
In addition, IC~443 is nearby at a distance of 1.5 kpc, compared to W44 (3~kpc) and W28 (2.5~kpc), which improves the IC~443 image linear resolution at the source (Fesen 1984; Mufson et al.\ 1986; Velazquez et al.\ 2002; Green 1989; Radhakrishnan et al.\ 1972; Kaspi et al.\ 1993; Hartl et al.\ 1983).
The shocked $1720\ {\rm MHz}$ OH emission in IC~443 was first detected by DeNoyer (1979a).

This paper presents new observations of the OH (1720~MHz) masers in IC~443 made using the Multi-element Radio Linked Interferometry Network\footnote{MERLIN is operated as a National Facility by the University of Manchester, Jodrell Bank Observatory, on behalf of Particle Physics and Astronomy Research Council (PPARC).} (MERLIN) and the Very Long Baseline Array (VLBA) of the NRAO\footnote{The National Radio Astronomy Observatory (NRAO) is a facility of the National Science Foundation operated under a cooperative agreement by Associated Universities, Inc.}.
This paper also examines an NRAO Very Large Array (VLA) `A' configuration observation of the IC~443 masers (from 12 March 1994) that has already been discussed by C97.
This paper will use the source numbering convention of Claussen et al.\ (C97 Table~4).

\section{Observations and Data Reduction}

\subsection{MERLIN}

The MERLIN radio telescope of the Nuffield Radio Astronomy Laboratories observed the IC~443 OH masers at $1720\ {\rm MHz}$ on 2000 January 20
for $\sim 9$~hours.
Six antennas were used, including the Mark II at Jodrell Bank, the 32~m antenna at Cambridge, and the 25~m dishes at Knockin, Darnhall, Tabley, and Defford.
Both right-handed and left-handed circular polarizations were recorded.
The correlator produced 512 spectral channels across a 500~kHz passband which yielded a velocity channel spacing of 0.17~\kms\ and a velocity resolution of 0.20~\kms.
The visibilities were integrated for 8.0~s.
The field of view of the array for these parameters is larger than the ${\sim}{10\arcsec}$ over which the maser emission is distributed (C97).

The absolute amplitude calibration was set by observations of 3C286.
The bandpasses were calibrated using observations of 3C84.
The phases were calibrated by frequent observations of 0617+210 at 13~MHz bandwidth and transferred to the 500~kHz maser bandwidth using the 3C84
offset solutions.
The {\it rms} noise in the final images is 11~\mjb, in agreement with expected values.
The synthesized beam of the images is $250\times155$~mas at a position angle $27{\arcdeg}$.

\subsection{VLBA+Y1}

We also observed the IC~443 OH (1720~MHz) masers on 2001 November 02 with the ten antennas of the VLBA plus one antenna of the VLA for
$\sim 5$~hours.
Both left and right circular polarizations were recorded.
The correlator produced 1024 spectral channels across a 1~MHz band yielding a 0.17~\kms\ velocity channel spacing and a velocity resolution of 0.20~\kms.
The visibilities were integrated for 8.9~s.
The field of view of the VLBA+Y1 array for these parameters is $15\arcsec$.
Approximately half of the data were unusable due to ionospheric and other effects.
In particular, the baselines to Mauna Kea did not have sufficient signal to allow self-calibration and were not used in the final images.
The synthesized beam of the resulting images is $15\times 12$~mas at a position angle of $12{\arcdeg}$.

The amplitude scale was set by online system temperature monitoring and {\it a priori} antenna gain measurements.
Bandpass responses and station delays were found from observations of J0555+3948.
The observations were phase referenced to J0557+2413 which is $4.8{\arcdeg}$ distant from the masers.
The phase reference source and the maser targets were alternately observed every 2~minutes to allow the maser visibility phases to be tracked against the brighter J0557+2413.
The {\it rms} noise in the final images is 25~\mjb, in agreement with expected values.

\section{Results}

\subsection{Images}

An image of the strongest maser channel ($V_{\rm LSR} = -4.64$~\kms) from the VLA data (C97) is shown in Figure~1a
(resolution $1.19\arcsec\times 0.92\arcsec$), while Figure~1b shows an image made from the combined VLA and MERLIN data for the same velocity
(resolution $590\times 540$~mas).
Figure~2 shows the ${V_{\rm LSR}}=-4.64$~\kms\ channel from the MERLIN data alone with a resolution of 250$\times$155~mas.
Three of the six features noted by C97 using the VLA are visible in the MERLIN image.
In addition, the MERLIN data has resolved the VLA Feature~1 into two peaks, 1A and 1B.
All four MERLIN sources are well resolved.
All image features were analyzed by fitting two-dimensional gaussian deconvolutions with the AIPS\footnote{The Astronomical Imaging Processing System (AIPS) software package of the NRAO is documented at {\tt http://www.nrao.edu/aips/}} task JMFIT.
The deconvolved angular sizes are $90 < \theta < 180$~mas.
Table~1 summarizes the MERLIN image features, including the brightness temperatures corresponding to the feature sizes.  Figure~2 also shows
the MERLIN Stokes $I$ line profiles from the maser peak positions.  The velocity widths and velocities of these features are summarized in
Table~3 and presented in \S3.2.  From Table~3, it is also clear that the MERLIN observations recover all of the line flux density observed for Features 1, 2, and 3 with
the VLA (C97).  Feature~5 is not detected in the MERLIN image (Fig.~2), although the fact that it is unresolved in the VLA data (Fig.~1a)
suggests that it should not be resolved out by MERLIN.  This discrepancy is most likely due to low level residual sidelobes that are apparent
in the MERLIN image (caused by insufficient UV coverage) at the position of Feature~5.  Since all of the other VLA line flux density is recovered in
the MERLIN image, we conclude that the masers do not exhibit structure on scales larger than $\sim200$~mas.  The 200~mas MERLIN angular
size upper limits correspond to $4\times 10^{15}$~cm (300~AU) at the distance of IC~443 (1.5 kpc).

Figure~3 shows the image of the $V_{\rm LSR} = -4.64$~\kms\ velocity channel from the VLBA+Y1 observations.  The three peaks lie within the
MERLIN 1A feature.  The VLBA was not able to detect the other, weaker MERLIN features.  The 1A core sizes are in the range $25 < \theta <
75$~mas, with brightness temperatures $(5-19) \times 10^8$~K, and are summarized in Table~2.  The VLBA image (Fig.~3) contains about 75\% of the line flux density in the MERLIN image of Feature~1A.
Thus we would argue that about 75\% of the maser emission comes from compact cores while about 25\% of the maser emission exists on
extended angular scales resolved out by the VLBA.  Figure~3 also shows the VLBA Stokes~$I$ spectra at the three image peaks whose properties are summarized in
Table~3.

\subsection{Line Widths}

The MERLIN and VLBA Stokes $I$ spectra at the peak maser positions were fitted with gaussian profiles using the GIPSY\footnote{The Groningen
Image Processing System (GIPSY) software package is documented at {\tt http://www.astro.rug.nl/${\mathtt \sim}$gipsy/}.}  software package.  The resulting
deconvolved line widths of the maser line profiles from the MERLIN and the VLBA observations are summarized in Table~3.  These data are
comparable to other OH (1720~MHz) maser observations, with typical line widths of $\sim 0.5$ \kms\/.  However, a few of the IC~443 SNR maser
lines are measured to have velocity widths less than the Doppler velocity width $\Delta{V}_D$ of the masing gas.
We estimate the minimum Doppler velocity width in the maser by assuming a gas temperature of $T = 90$~K in accord with the molecular studies of the shocked region ({\it e.g}.\ van~Dishoeck, Jansen, \& Phillips 1993) and by ignoring the Doppler contributions of any bulk turbulent flows in the shock.
By this method we find $\Delta V_D = 0.49$~\kms\ in IC~443 clump G1.
Any contributions from turbulent motions, which are expected to be small, would broaden the expected $\Delta V_D$ and further stress that the maser lines are sub-Doppler.
Thus, the MERLIN 1A line ($\Delta V = 0.42\pm0.05$~\kms) and the VLBA 1A1 line ($\Delta V = 0.24\pm0.07$~\kms) are appreciably narrower than the expected Doppler profile.
The significance of these sub-Doppler line widths is discussed in \S 4.2.

\subsection{Polarimetry}

Figure~4 shows the Stokes $I$ and $V$ profiles from the 1A source peak of the MERLIN image in Figure~2.
This figure demonstrates that the 1A maser is $\sim 10$\% right circularly polarized.
The Stokes $V$ profiles for MERLIN Features 1B, 2, and 3 and VLBA Feature~1A1 also show $\sim 10$\% circular polarization.
Features 1B and 2 are right circularly polarized, while Feature~3 is left circularly polarized.
The Stokes $V$ profiles from the C97 VLA data are in agreement with these MERLIN and VLBA results, in both the magnitude and sense of circular polarization.
These IC~443 circular polarization results are in distinct contrast to the Zeeman {\sf S}-shaped profiles typically observed in OH (1720~MHz) SNR masers ({\it e.g}.\ Brogan et al.\ 2000; C99; C97). 
 
\section{Discussion}

\subsection{Sizes}

The geometry at the shock front is an important consideration when discussing the measured maser sizes.
OH (1720~MHz) SNR masers are observed only at the SNR/molecular cloud interface where the shock front is moving transversely across the plane of the sky ({\it e.g}.\ Frail et al.\ 1996).
IC~443 provides an excellent example of this geometrical selection effect.
There is a $\sim 9$~pc ring of shocked molecular interaction around the SNR ({\it e.g}.\ Dickman et al.\ 1992) but the masers are only observed in clump G where the shock is moving transverse to the line of sight ({\it e.g}.\ Wang \& Scoville 1992).
This transverse motion allows maximum velocity coherence along the line of sight.
The post-shock region that provides the collisional pump conditions (LGE; Wardle 1999) is determined by the downstream temperature profile needed for the collisional excitation and the dissociative SNR X-rays that generate OH molecules from H$_2$O that is formed upstream.

LGE suggest that the size of the region suitable for population inversion and maser emission $L$ is ${\sim}10^{16}$~cm (${\sim}700$~AU).
This transverse scale $L$ is much smaller than the measured size of the shocked clump G region:  $(5-8) \times 10^{17}$~cm (Turner et al.\
1992; van~Dishoeck et al.\ 1993; Tauber et al.\ 1994; Cesarsky et al.\ 1999).  However, these sizes are only an upper limit to the possible
transverse sizes $\ell_m$ of the masers.  The physical conditions required for the maser emission are more stringent than those responsible
for the shocked emission observed in the molecular studies.  We measure angular sizes in the range $90 < \theta < 180$~mas, corresponding to
linear sizes $135\ {\rm AU}<{\ell_m}<270$~AU.  Therefore, less than 10\% of the transverse width $L$ of the IC~443 clump G region that can theoretically
produce stimulated emission shows maser radiation.

The multiple compact cores observed in IC~443 with the VLBA (Fig.~3) are not unique in the OH (1720~MHz) SNR maser class.  A similar
core/diffuse morphology is observed using MERLIN and the VLBA by C99 in the W44 and W28 masers.  They also note that
MERLIN recovers all of the line flux density observed with the VLA.  The VLBA observation of the W44 E11 maser source shows a compact core
responsible for about half of the emission.  All of the flux density of the W28 E24, E30, and E31 maser sources is resolved out completely in
the VLBA data.  In fact, all of the W28 and W44 masers imaged with the VLBA by C99 show some $\sim 100$~mas structure to
which MERLIN is sensitive, but which is resolved out by the VLBA observations.  Since the intrinsic structures in the IC~443, W28, and W44
masers are comparable, this morphology likely indicates further physical similarity among all OH (1720~MHz) SNR maser regions.

We assume that the line of sight to IC~443 is relatively free of scatter-broadening effects ({\it e.g}.\  Lazio \& Cordes 1998a,b). We have
also measured linear sizes comparable to those observed on MERLIN and VLBA scales by C99 in W28 and W44.  These angular size and morphology
measurements indicate that all MERLIN and VLBA OH (1720~MHz) angular size measurements to date ({\it e.g}.\ C99; Brogan et al.\ 2002), except
those in the Galactic center (Yusef-Zadeh et al.\ 1996), have determined the intrinsic maser sizes.  In general it appears that a ${\sim}$300~AU region of
diffuse maser emission surrounds multiple compact cores ${\sim}$60~AU in transverse size.

\subsection{Line Widths}

Using the narrow line widths of the MERLIN 1A line and the VLBA 1A1 line (Table~3), it is possible to place constraints on the geometry of
the maser regions.
Elitzur (1998) suggests that the Doppler width $\Delta V_D$ expected from the line-emitting gas will be related to the emitted maser line width $\Delta V$ by $\Delta V = \Delta V_D / \sqrt{p}$ such that integer values of $p = 1, 2, 3$ correspond to the dimension of filamentary, planar, or ellipsoidal region geometries, respectively.
Comparison of this relation with the velocity widths for Features~1A (MERLIN) and 1A1 (VLBA) in Table~3, implies that at least some of the masers are emitted from regions with $p\approx3$.
The gas region geometry implied for these features by this line narrowing is roughly ellipsoidal; the data are not consistent with a thin emitting filament along the line of sight.
LGE find that an OH (1720~MHz) maser region with an aspect ratio of three can yield a maser brightness temperature $T_B < 10^9$~K.
This prediction is in good agreement with the brightness temperatures listed in Tables~1 and 2.

We will not discuss various maser line `rebroadening' theories ({\it e.g}.\ Litvak 1970; Peters \& Allen 1972) that depend explicitly on saturation because we have been unable to constrain the saturation state of the masers.
Elitzur (1998) suggests a test for maser saturation that requires an antisymmetric Stokes $V$ line profile but, as discussed in \S4.3, we do not observe an antisymmetric Stokes $V$ line profile in the IC~443 masers.
Since we may only assume that the masers are saturated on the grounds that they have not been observed to vary (see \S\S3.1, 4.4), the only warranted line width discussion is the geometry theory discussed above.

\subsection{Polarization}

The Zeeman splitting of the OH (1720~MHz) SNR maser line emission from supernova remnants has proven to be an outstanding tool for measuring
the magnetic field strengths in supernova shock fronts ({\it e.g}.\ Brogan et al.\ 2000).
The maser Stokes~$V$ line profiles often show the expected antisymmetrical {\sf S}-shaped thermal Zeeman pattern.
In some cases, spatial or spectral blending of different maser components partially confuse the expected Zeeman profile.
However, IC~443 represents the first observation of single-handed circularly polarized line spectra from OH (1720~MHz) masers in supernova remnants (see Fig.~4).
There are a number of effects which could potentially cause the observed circular polarization, and many of these could act simultaneously:  (1) spatial blending of masing spots; (2) spectral blending of individual maser lines; (3) velocity gradients; (4) magnetic field gradients; and (5) Faraday rotation.
In the following, we are able to rule out (1) and (2) from our observational results.
We cannot constrain the contributions of (3), (4), and (5) but we give a few examples of how these effects could conspire to yield the observed polarization.

In general, if spatially distinct maser features emit at different velocities or different luminosities, their emission would not present a
symmetrical {\sf S}-shaped Zeeman or gaussian Stokes $I$ profile when convolved together in a large beam.  For example, different OH
(1720~MHz) SNR maser source spectra are spatially blended in VLA observations of W28 and W44 (C97) and do not present simple Zeeman Stokes
$V$ profiles.  However, subsequent MERLIN and VLBA observations toward W28 and W44 do discern simple Zeeman spectra from multiple
spatial components (C99).  Similarly, spatial blending is a complicating factor for the interpretation of IC~443 Feature~1 (Fig.~1).  VLA
observations (C97) show a single maser feature while MERLIN observations (Fig.~2) reveal the presence two resolved maser peaks, 1A and 1B.
The MERLIN 1A maser feature, in turn, has emission from at least three distinct sources (1A1, 1A2, and 1A3; Fig.~3) on VLBA size scales.
Surprisingly, the VLA, MERLIN, and VLBA data, covering three orders of magnitude in angular resolution, all show the same ${\sim}$10\%
circular polarization.  Thus, although these IC~443 VLBA observations have the highest linear resolution of any OH (1720~MHz) SNR maser study
to date, no symmetrical {\sf S}-shaped Zeeman line profiles are observed to be spatially blended.  Therefore, unlike the W28 and W44 masers
for which subsequent MERLIN and VLBA observations do discern simple Zeeman spectra (C99), we observe no evidence that spatial blending is
responsible for confusing the circular polarization signal in the IC~443 OH (1720~MHz) masers.

Spectral blending may explain the circular polarization levels even when spatial blending cannot.  Spectral blending, however, is not
discernible with multiple angular resolution studies since all of the emission components lie along the line of sight.  It is possible that
several different masers, each with the expected Zeeman profile, lie along the line of sight with different velocities, line widths, and
amplitudes.  In specific combinations these masers could be superposed to yield a 10\% circularly polarized line profile.  However, it is
unlikely that all of the spatially distinct lines-of-sight in the IC~443 maser images (Figs.~1, 2, \& 3) contain these exact superpositions necessary to reproduce the $\sim 10$\% circular polarization levels.
Additionally, the sub-Doppler line widths observed in the IC~443 masers (\S 3.2, 4.2) limit an explanation
involving spectral blending, since several line components at different velocities would be required, and such blending must necessarily
broaden the observed Stokes $I$ profile that is their sum.  For these reasons, we do not believe that spectral blending can account for the
symmetrical Stokes $V$ profiles observed for the OH (1720 MHz) masers in IC~443.  

Since it appears that neither spectral blending nor spatial blending significantly affects the line profiles observed in the IC~443 OH (1720~MHz) SNR masers, the observed ${\sim}$10\% circular polarization levels are most likely an intrinsic property of the maser emission.
Elitzur (1992) reviews a number of polarization selection mechanisms that could affect the OH (1720~MHz) maser collisional pump, including velocity or magnetic field gradients.
For example, Cook (1966) and Shklovskii (1969) require both a magnetic field and velocity gradient along the line of sight in order to amplify one hand of circular polarization.
Similarly, Deguchi \& Watson (1986) predict a circularly polarized line in the presence of a uniform magnetic field and a large velocity gradient.
A line-of-sight magnetic field ``beaming'' geometry has also been suggested to affect the degree of circular polarization (Gray \& Field 1994).
Alternatively, Deguchi \& Watson (1993) suggest that if the Faraday rotation within the gain medium is significant, Stokes $U$ polarization can be transfered to Stokes $V$ signal.
For this Stokes transfer mechanism, Field \& Gray (1994) estimate that an electron column density of $N_e > 10^{17}\ {\rm cm}^{-2}$ is required.
The studies of IC~443 clump G by Bykov et al.\ (2000), Turner et al.\ (1992), and Keohane et al.\ (1997) indicate electron column densities of this order.
However, it is not clear how the OH (1720~MHz) maser polarization data for IC~443 -- (1) the 10\% circular polarization (Fig.~4) and (2) the 4.2\% linear polarization (VLA: C97) -- can apply or constrain these theories.

\subsection{Proper Motion}

Dickman et al.\ (1992) consider it plausible that clump~G was a preexisting molecular cloud condensation, only now being encountered by the SNR blast wave.
Theories due to McKee et al.\ (1978) and, if applicable, Bertoldi \& McKee (1990) suggest that a cloud with the size and mass of clump G should survive the shock ram and be accelerated into ballistic motion.
Indeed, Wang \& Scoville (1992) find IC~443 clumps C1 and C2 to have been accelerated in this way.
Clump G may be expected to have a ballistic velocity on the order of the shock velocity $v_s \approx 40$~\kms\ ({\it e.g}.\ van~Dishoeck et al.\ 1993; Draine, Roberge, \& Dalgarno 1983).
Thus, the phase-referenced VLBA maser observations of IC~443 (Fig.~3) are suitable as the first epoch of a possible $\sim 30$~mas proper motion measurement over $\sim 5$~yr.

Alternatively, clump G may remain stationary.
In such a case, the maser lifetime can be expected to be determined by the relatively short shock crossing time $L/v_s \approx 10^3$~yr.
Indeed, Cesarsky et al.\ (1999) finds that clump G is not in steady state and Rho et al.\ (2001) suggest the presence of a time-variable $C$-type shock to explain their [O{\sc i}] 63.18~$\mu$m line and infrared continuum data.
However, our data do not show any time variability in the maser emission over the six years between the VLA and MERLIN observations (see \S3.1).

Furthermore,  McKee \& Tan (2002) suggest that stars may form in $\sim 10^5$~yr in physical conditions satisfied by OH (1720~MHz) SNR maser regions such as IC~443 clump~G and W44 (cf.\ Plume et al.\ 1997; Wootten 1978).
Indeed, star formation theory often identifies shock fronts as the agent by which molecular clouds are compressed to high density ($n > 10^5\,{\rm cm}^{-3}$) and triggered into the gravitational collapse of new stars ({\it e.g}.\ Elmegreen \& Lada 1977).
However, Dickman et al.\ (1992) expect the clumps to retain their ``kinematic signatures'' for at least $\sim 10^6$~yr.

The timescales of the various dynamical predictions applicable to IC~443 clump G span three or four orders of magnitude.
A proper motion study of the masers in these environments may distinguish between the different theoretical scenarios.
Also, future study of the IC~443 system could considerably illuminate the degree to which OH (1720~MHz) SNR masers mark the sites of future star formation in molecular clouds ({\it e.g}.\ Wardle \& Yusef-Zadeh 2002).

\section{Conclusions}

Interstellar image scatter-broadening is expected to be minimal toward IC~443 near the Galactic anticenter.
The MERLIN and VLBA linear resolution at the relatively close 1.5~kpc distance to the supernova remnant is the most favorable of any investigation to date.
Thus, these data represent the most reliable measurement of the intrinsic sizes of OH (1720~MHz) masers in supernova remnants.
Also, we observe multiple compact maser cores at VLBA angular scales, as has been observed in other VLBI studies of OH (1720~MHz) masers ({\it e.g}.\ C99).
If we generalize these findings to all similar masers, these results indicate that OH (1720~MHz) SNR masers have a core/diffuse morphology with linear sizes approaching ${\sim}300$~AU and cores smaller than ${\sim}$60~AU.
These measured sizes also indicate that approximately 10\% of the shocked ``clump G'' region in IC~443 capable of supporting the maser actually shows maser emission.

The circular polarization profiles of the IC~443 OH masers are unique in their class.
We also present the narrowest line observed from OH (1720~MHz) SNR masers.
Future work on IC~443 masers should include high spectral and spatial resolution linear polarization imaging.
Linear polarization information may help constrain the role of the magnetic field in both the Zeeman behavior and the geometry of the masing gas ({\it e.g}.\ Elitzur 1998).
A second epoch of phase-referenced proper motion observations of the IC~443 masers may yield information on the dynamical lifetime of the maser regions.
In conclusion, though IC~443 contains a relatively small number of OH (1720~MHz) masers compared to other SNRs, IC~443 is an excellent laboratory for the study of the polarization and morphology properties that seem to be ubiquitous in the OH (1720~MHz) SNR maser class.

\acknowledgments

We wish to thank M. Elitzur for helpful comments on an earlier draft.
We also thank the referee for timely correspondence.
IMH was supported in part by the NRAO summer student program and the NRAO pre-doctoral researcher program.

\clearpage

\begin{figure}
\plotone{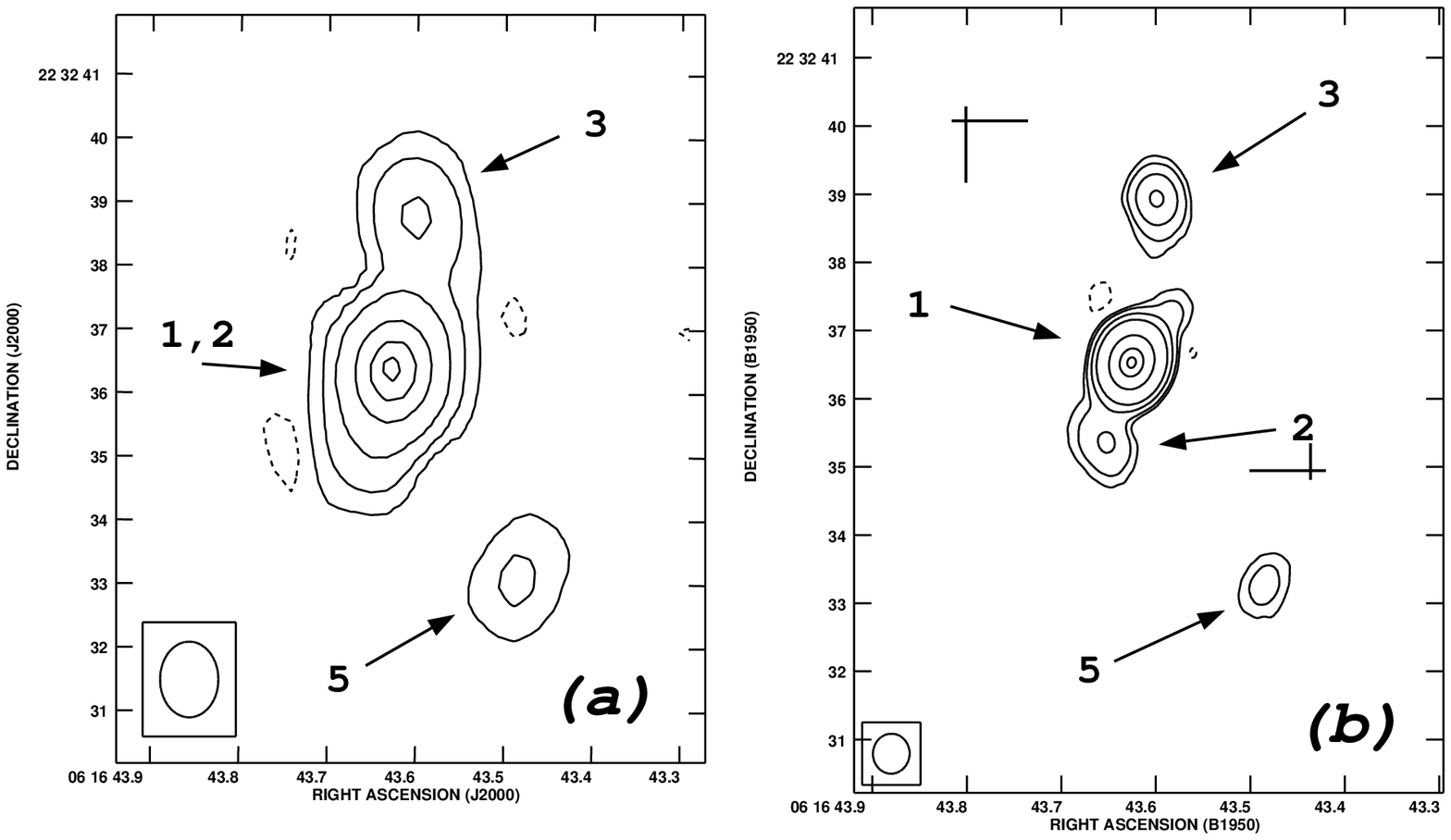}
\caption{VLA and MERLIN+VLA images at $V_{\rm LSR} = -4.64\,{\rm km}\,{\rm s}^{-1}$.
{\it (a)}- VLA A array image of IC~443 masers from Claussen et al.\ (C97).
The contours are -3, 5, 20, 75, 300, 600, and 825 times the image {\it rms} $4.3\ {\rm mJy}\ {\rm beam}^{-1}$.
The VLA observations have a velocity resolution of 0.53~\kms.
The beam is $1.19" \times 0.92"$ at a position angle of $0\arcdeg$.
{\it (b)}- Image of combined MERLIN and VLA data as described in the text.
The contours are -5, 8, 15, 35, 75, 300, 600, and 750 times the image {\it rms} $4.3\ {\rm mJy}\ {\rm beam}^{-1}$.
The beam is 590$\times$540~mas at a position angle of $-3\arcdeg$.
The perpendicular lines in Figure~1b indicate the corners of the MERLIN image in Figure~2.
\label{fig1}}
\end{figure}

\clearpage

\begin{figure}
\plotone{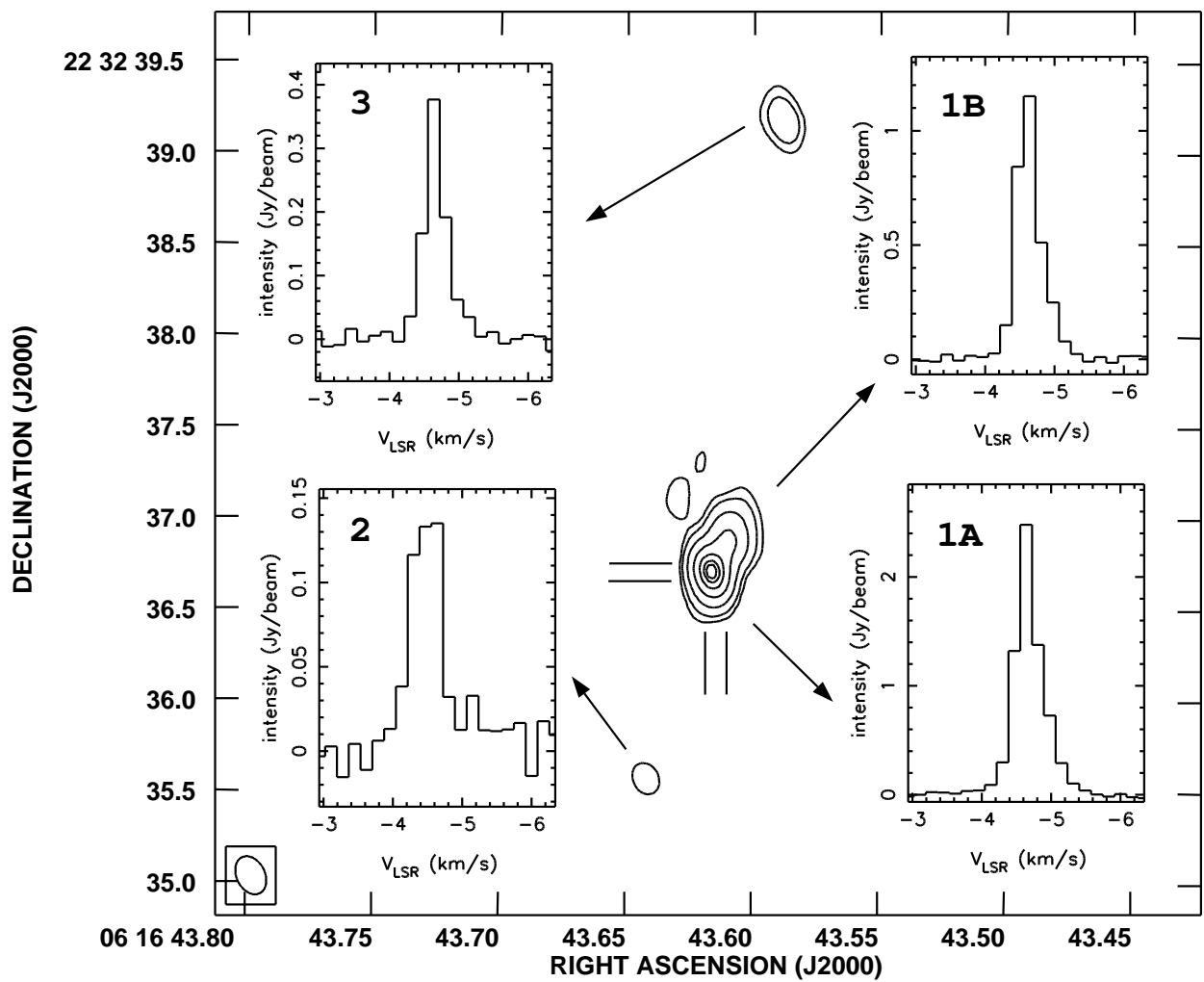}
\caption{MERLIN image of IC~443 masers at $V_{\rm LSR} = -4.64\,{\rm km}\,{\rm s}^{-1}$.
The contours are -5, 8, 16, 40, 90, 150, 190, and 210 times the image {\it rms} $11\ {\rm mJy}\ {\rm beam}^{-1}$.
The beam is 250$\times$155~mas at a position angle of $27\arcdeg$.
The spectra are Stokes $I$ profiles at the peaks indicated by the arrows.
The spectral resolution is 0.20~\kms.
Line profile properties are summarized in Table~3.
The straight lines near the central peak indicate the extended edges of the VLBA image in Figure~3.
\label{fig2}}
\end{figure}

\clearpage

\begin{figure}
\plotone{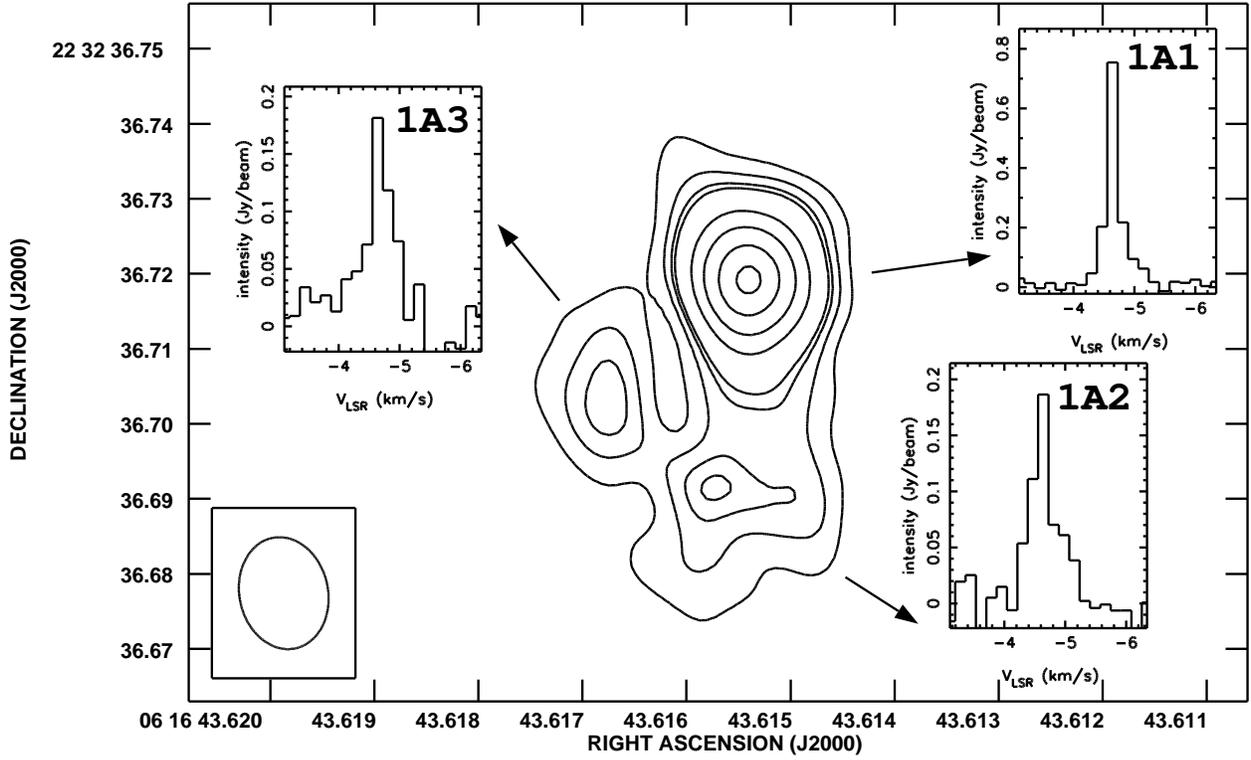}
\caption{VLBA image of the IC~443 compact maser cores at $V_{\rm LSR} = -4.64\,{\rm km}\,{\rm s}^{-1}$.
The contours are -5, 3.5, 5, 6.25, 7, 12, 17, 23, and 28 times the image {\it rms} $26\ {\rm mJy}\ {\rm beam}^{-1}$.
The beam is 15$\times$12~mas at a position angle of $12\arcdeg$.
All three peaks lie within the MERLIN 1A image feature (Fig.~2).
The spectra are Stokes $I$ profiles taken at the peaks indicated by the arrows.
The velocity resolution is 0.20~\kms.
Line profile properties are summarized in Table~3.
\label{fig3}}
\end{figure}

\clearpage

\begin{figure}
\plotone{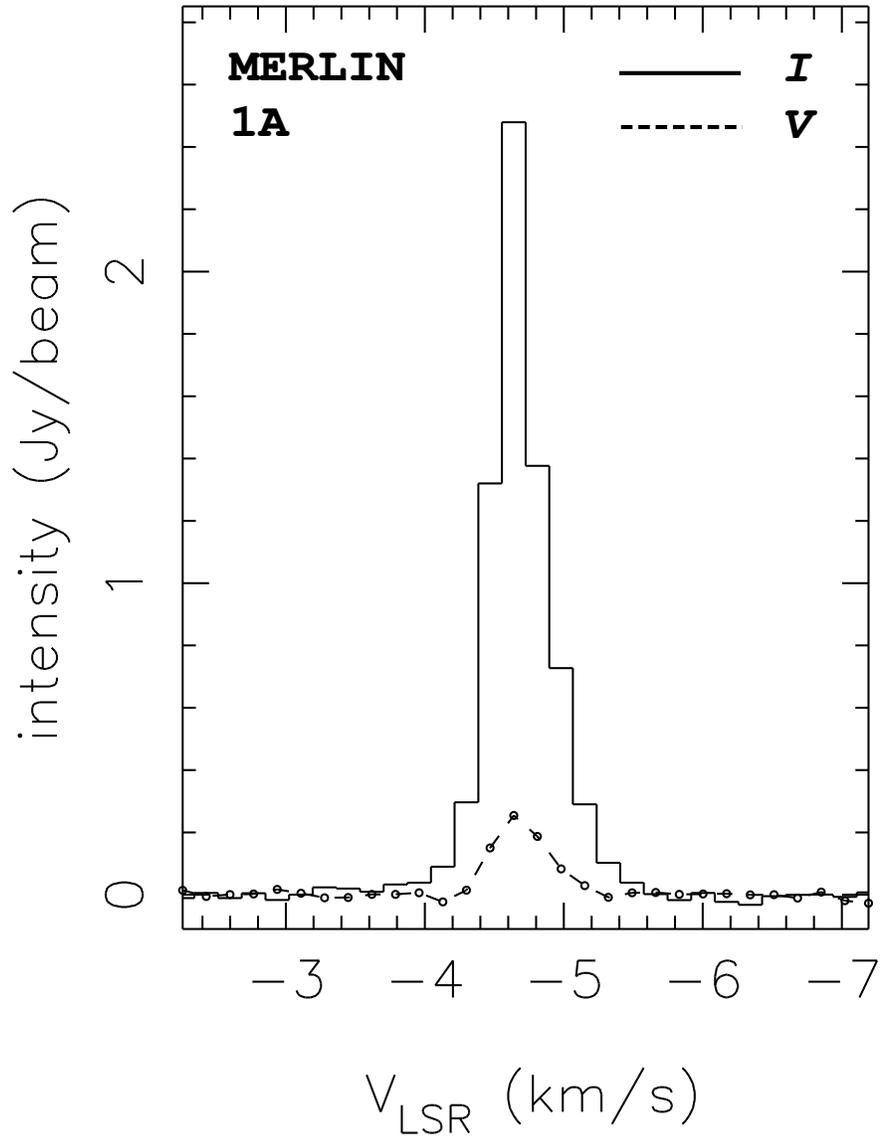}
\caption{MERLIN Stokes $I$ and $V$ line profile comparison.
These spectra for MERLIN image Feature~1A (Fig.~2).
Intensity (Jy/beam) versus $V_{LSR}$ (km/s).
For these spectra $V \approx 0.1 I$.
All of the maser lines in IC~443 show similar ${\sim}$10\% circularly polarized profiles rather than antisymmetrical {\sf S}-shaped Stokes $V$ Zeeman detections as observed in many OH (1720~MHz) SNR maser sources.
\label{fig4}}
\end{figure}

\clearpage

\begin{deluxetable}{ c r r c c c }
\tabletypesize{\scriptsize}
\tablecaption{MERLIN Image Features \label{tbl-1}}
\tablewidth{0pt}
\tablehead{
\colhead{Feature} & \colhead{R.A.\tablenotemark{a}} & \colhead{Dec.\tablenotemark{a}} & \colhead{$I$} & \colhead{$\theta_{\rm max}$} & \colhead{$T_B$} \\
 & \colhead{(h m s)} & \colhead{($^\circ$ $'$ $''$)} & \colhead{(\mjb)} & \colhead{(mas)} & \colhead{($10^6$~K)} \\
}
\startdata
 3 & 06 16 43.588(2) & 22 32 39.18(1) &  377(11) & 110(10) & 5.2(5) \\
1B &       43.610(2) &       36.89(1) & 1151(11) & 180(10) & 13(1) \\
1A &       43.615(2) &       36.72(1) & 2480(11) & 120(10) & 34(3) \\
 2 &       43.641(2) &       35.57(1) &  135(11) &  90(10) & 1.8(2) \\
\enddata
\tablenotetext{a}{Coordinates in J2000 epoch}
\tablecomments{These positions differ from C97 by ${\sim}$40~mas due to updated VLBI positions for the phase calibrators 0629+104 and 0617+210.}
\end{deluxetable}

\clearpage

\begin{deluxetable}{ c r r c c c }
\tabletypesize{\scriptsize}
\tablecaption{VLBA Image Features \label{tbl-2}}
\tablewidth{0pt}
\tablehead{
\colhead{Feature} & \colhead{R.A.\tablenotemark{a}} & \colhead{Dec.\tablenotemark{a}} & \colhead{$I$} & \colhead{$\theta_{\rm max}$} & \colhead{$T_B$} \\
 & \colhead{(h m s)} & \colhead{($^\circ$ $'$ $''$)} & \colhead{(\mjb)} & \colhead{(mas)} & \colhead{($10^8$~K)} \\
}
\startdata
1A1   & 06 16 43.6154(3) & 22 32 36.719(2) & 750(25) & 15(3) & 19(4) \\
1A2   &       43.6157(8) &       36.691(3) & 190(25) & 55(4) & 4.9(4) \\
1A3   &       43.6167(8) &       36.703(3) & 180(25) & 25(4) & 4.7(4) \\
\enddata
\tablenotetext{a}{Coordinates in J2000 epoch}
\tablecomments{These positions differ from C97 by ${\sim}$40~mas due to updated VLBI positions for the phase calibrators 0629+104 and 0617+210.}
\end{deluxetable}

\clearpage

\begin{deluxetable}{ r r | r r r r | r r r r } 
\tabletypesize{\scriptsize}
\tablecaption{Maser Flux Density and Line Profiles \label{tbl-3}}
\tablewidth{0pt}
\tablehead{
\multicolumn{2}{c}{VLA} & \multicolumn{4}{c}{MERLIN} & \multicolumn{4}{c}{VLBA} \\
\colhead{Feature} & \colhead{$S$} &
\colhead{Feature} & \colhead{$S$} & \colhead{$V_{\rm LSR}$} & \colhead{$\Delta{V}$} &
\colhead{Feature} & \colhead{$S$} & \colhead{$V_{\rm LSR}$} & \colhead{$\Delta{V}$} \\
 & \colhead{(mJy)} & & \colhead{(mJy)} & \colhead {(\kms)} & \colhead {(\kms)} & & \colhead{(mJy)} & \colhead {(\kms)} & \colhead {(\kms)} \\
}
\startdata
1 & 4180(50) & 1A & 2880(60) & -4.60(2) & 0.42(5) & 1A1 & 1020(30) & -4.60(3) & 0.24(7) \\
  &          &    &          &          &         & 1A2 &  370(30) & -4.70(4) & 0.41(9) \\
  &          &    &          &          &         & 1A3 &  280(30) & -4.70(4) & 0.46(9) \\
  &          & 1B & 1270(60) & -4.60(2) & 0.48(5) & \tablenotemark{a} & & & \\ \tableline
2 &  180(50) &  2 &  150(20) & -4.50(4) & 0.63(9) & \tablenotemark{a} & & & \\ \tableline
3 &  390(20) &  3 &  390(20) & -4.60(2) & 0.52(5) & \tablenotemark{a} & & & \\ \tableline
5 &  110(20) & \tablenotemark{a} & & & & \tablenotemark{a} & & & \\
\enddata
\tablenotetext{a}{feature not detected}
\end{deluxetable}

\end{document}